\documentclass[preprint,12pt]{elsarticle}

\usepackage{graphicx,
dcolumn,
amsmath,
amssymb,
color,
mathrsfs,
titlesec,
hyperref,
enumerate,
bm,
subfigure,
float,
comment,
lineno,
soul,
xspace,
xcolor,
}
\usepackage[export]{adjustbox}

\usepackage{todonotes}

\biboptions{sort&compress}

\newcommand{\me}{\mathrm{e}}
\newcommand{\mi}{\mathrm{i}}
\renewcommand{\vec}[1]{\bm{#1}}
\newcommand{\smat}{\ensuremath{\mathcal{S}}-matrix\xspace}
\newcommand{\smats}{\ensuremath{\mathcal{S}}-matrices\xspace}

\newcommand{\revise}[1]{{#1}} 

\emergencystretch=\maxdimen
\usepackage{hyperref}
\definecolor{redlinks}{rgb}{1,0,0}
\hypersetup{
  colorlinks=true,
}

\journal{Ultramicroscopy}

\begin{document}

\begin{frontmatter}
\title{Uncovering polar vortex structures by inversion of multiple scattering with a stacked Bloch wave model}

\author[UCBMSE]{Steven E Zeltmann}
\ead{steven.zeltmann@berkeley.edu }

\author[MSD]{Shang-Lin Hsu}

\author[melb]{Hamish G. Brown}

\author[ASU]{Sandhya Susarla}

\author[UCBMSE]{Ramamoorthy Ramesh}
\author[UCBMSE,NCEM]{Andrew M Minor}
\author[NCEM]{Colin Ophus}
\ead{cophus@gmail.com}

\address[UCBMSE]{Department of Materials Science and Engineering, University of California, Berkeley, Berkeley, CA}
\address[MSD]{Materials Science Division, Lawrence Berkeley National Laboratory, Berkeley, CA}
\address[melb]{Ian Holmes Imaging Centre, Bio21 Molecular Science and Biotechnology Institute, University of Melbourne, Victoria, Australia}
\address[ASU]{School for Engineering of Matter, Transport and Energy, Arizona State University, Tempe, AZ}
\address[NCEM]{National Center for Electron Microscopy, Molecular Foundry, Lawrence Berkeley National Laboratory, Berkeley, CA}

\begin{abstract}

Nanobeam electron diffraction can probe local structural properties of complex crystalline materials including phase, orientation, tilt, strain, and polarization. Ideally, each diffraction pattern from a projected area of a few unit cells would \revise{produce a clear Bragg diffraction pattern}, where the reciprocal lattice vectors can be measured from the spacing of the diffracted spots, and the spot intensities are equal to the square of the structure factor amplitudes. However, many samples are too thick for this simple interpretation of their diffraction patterns, as multiple scattering of the electron beam can produce a highly nonlinear relationship between the spot intensities and the underlying structure. Here, we develop a stacked Bloch wave method to model the diffracted intensities from thick samples with structure that varies along the electron beam. Our method reduces the large parameter space of electron scattering to just a few structural variables per probe position, making it fast enough to apply to very large fields of view. We apply our method to SrTiO$_3$/PbTiO$_3$/SrTiO$_3$ multilayer samples, and successfully disentangle specimen tilt from the mean polarization of the PbTiO$_3$ layers. We elucidate the structure of complex vortex topologies in the PbTiO$_3$ layers, demonstrating the promise of our method to extract material properties from thick samples.

\end{abstract}


\end{frontmatter}


\hypersetup{
    linkcolor=redlinks,
    urlcolor=redlinks,
    citecolor=redlinks
}


\section{Introduction}
\label{Sec:Intro}

Multiple scattering is often viewed as an unwanted and cumbersome artifact in electron microscopy as it is responsible for confounding effects such as contrast reversals in phase contrast images and electric field maps \cite{yang2017electron, seki2022linear}, complex features inside nanobeam diffraction disks that hinder precise strain mapping \cite{mahr2015theoretical, cooper2016strain, pekin2017strain, zeltmann2020patterned}, and failure of many super-resolution imaging techniques \cite{chen2021electron}. 
\revise{Despite this, the rich physics of dynamical diffraction also provides us with the ability to very precisely measure material properties. 
John Spence pioneered the practice of inverting the dynamical scattering process to retrieve the features of materials, developing numerous algorithms for recovering detailed structural information from electron diffraction patterns \cite{PhysRevLett.125.065502,spence1994minimum,spence1998direct,weierstall2002image}.
Following Spence's legacy, this contribution follows applies the ideas of dynamical inversion to a large-area four-dimensional scanning transmission electron microscopy dataset.
}

\begin{figure}[tbp]
    \centering
    \centerline{\includegraphics[width=0.6\textwidth]{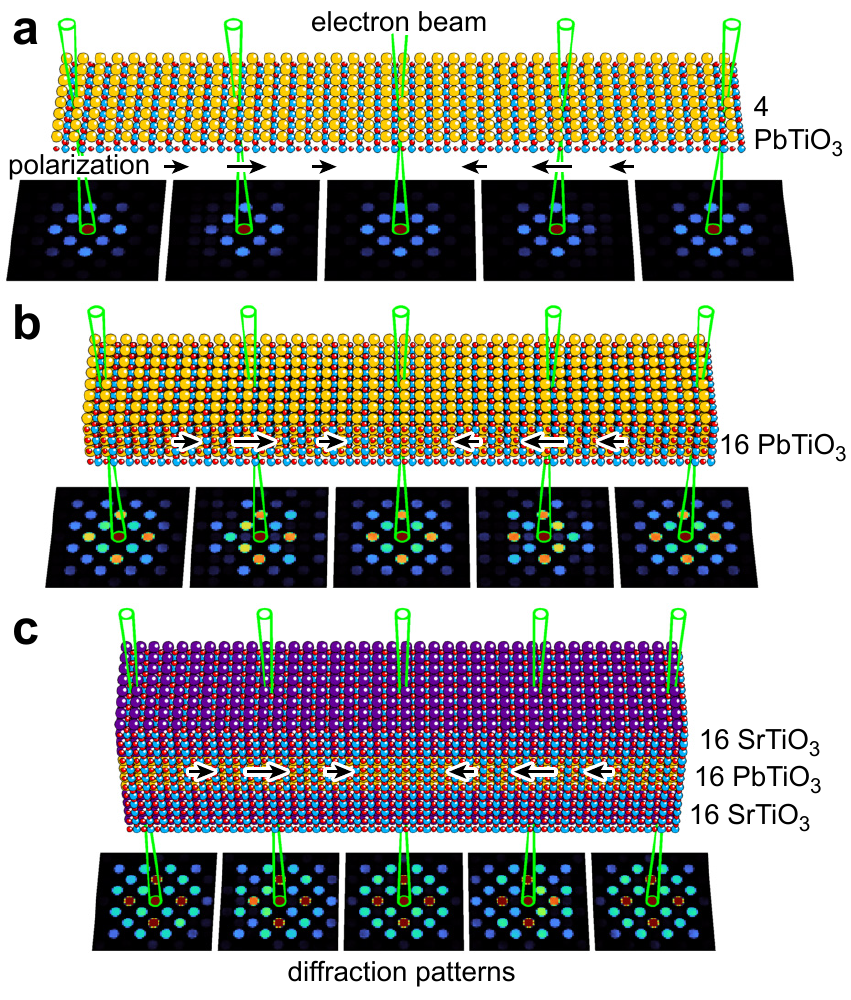}}
    \caption{{Nanobeam electron diffraction signals from lattices with varying in-plane polarization, indicated by the arrows.} Diffraction pattern simulations of (a) thin \revise{(4 u.c.)} PbTiO$_3$, (b) thick \revise{(16 u.c.)} PbTiO$_3$, and (c) multilayer with 16:16:16 unit cells of SrTiO$_3$:PbTiO$_3$:SrTiO$_3$. Left to right, the in-plane PbTiO$_3$ polarization varies smoothly from zero, full left-facing, zero, full right-facing, and zero polarization.}
    \label{fig:schematic}
\end{figure}

In a na\"ive sense, an electric field built in to a thin sample causes a tilt of the electron wave as it propagates through the material. 
Depending on the optical setup and the length scale of the changes in the field this manifests as an intensity redistribution in a diffraction pattern when using a larger convergence angle \cite{muller2012strain}, or a shift of the diffraction pattern when using a small convergence angle \cite{clark2018probing,cao2018theory}. 
In materials where the polarization is associated with a structural distortion, the diffracted intensities also change as a result of the change in the lattice electrostatic potential \cite{das2019observation}. 
This is shown schematically in Fig.~\ref{fig:schematic}a, which shows a sequence of diffraction patterns simulated at differing polarization for a thin \revise{(4 u.c., $\approx 1.6$~nm)} sample of PbTiO$_3$ (PTO).

When the sample thickness increases and multiple scattering occurs, however, the changes in the diffraction patterns become far more complex than this description.
Deb et al.~showed that when diffraction disks overlap, under the weak phase approximation there should be no contrast between opposing pairs of diffraction disks due to polarization; however, anomalous contrast between Friedel pairs arises when multiple scattering pathways are considered \cite{deb2020imaging}. 
Mahr et al \cite{MAHR2022113503} showed that for an interface with an electric field due to a difference in mean inner potentials, dynamical scattering causes the measured electric field to oscillate wildly for most experimental setups, with beam precession providing the only partial remedy. 
Nguyen et al \cite{nguyen2020transferring} also observed that when measuring chiral polarization domains via the diffraction intensity changes associated with a structural distortion, the chiral directions flip as a function of thickness. 
This case is shown schematically in Fig.~\ref{fig:schematic}b for sample of PTO that is thick enough \revise{(16 u.c., $\approx 6.3$~nm)} to cause multiple scattering of the electron probe.

Various approaches for reconstructing sample properties or structure under conditions of multiple scattering have been developed that utilize the \smat description of the scattering process \cite{brown2018structure, PhysRevLett.125.065502, pelz2021phase, findlay2021scattering, brown2022three}. 
In these approaches, the electron scattering process is encapsulated in the ``scattering matrix,'' an object which contains the information about the material structure and thickness-propagation effects, such that the \smat multiplies with a vector representing the incident electron wave to yield a vector representing the scattered wave \cite{sturkey1962calculation}. 
As we will discuss in the theory section below, there is useful correspondence between the entries of the \smat and the properties of the sample, such as its thickness, polarization, or tilt.
This correspondence can be used in both the forward direction to simulate diffraction intensities given known structural parameters, as well as in the reverse direction to recover structural parameters from diffraction intensities. 
Much of the literature on this method is concerned with atomic-resolution reconstruction of the sample potential.
Donatelli and Spence demonstrated a method for recovering the sample potential at high resolution from a tilt series of diffraction patterns by iterative inversion of the \smat and without prior knowledge of the sample thickness \cite{PhysRevLett.125.065502}.
Brown and coworkers have developed methods for recovery of the \smat, originally from focal series 4D-STEM data \cite{brown2018structure,brown2022three} and later extended to a single defocus \cite{findlay2021scattering}. Their former method provided some 3-dimensional information about the sample, which they recovered from the \smat by an optical sectioning approach. 
Alternatively, when the approximate structure of the material is known, and variation along the beam direction can be ascribed to a small number of parameters, the PRIMES (parameter retrieval and inversion from multiple electron scattering) family of methods can be used to obtain property variation along the beam direction \cite{PhysRevB.89.205409,PENNINGTON2015105,PENNINGTON201542,PhysRevB.97.024112}. 
These methods all utilize a stacked \smat model to represent parameter variation along the beam direction, which are refined against test CBED patterns using various numerical optimization schemes. 
\revise{Similarly, Jacob et al used quantitative CBED (QCBED) for the determination of structural information from buried interfaces by modeling of the dynamical scattering \cite{Jacob2008}. }

In this work, we analyze a complex sample consisting of multiple distinct layers through the thickness, and use a model of the electron multiple scattering to extract information about the material. The sample, consisting of 16 unit cells of SrTiO$_3$ (STO), 16 unit cells of PTO, and 16 units cells of STO \revise{(total thickness $\approx 19$~nm)}, is shown in Fig~\ref{fig:schematic}c.
The vortex structures in this material have been previously studied by plan-view and cross section transmission electron microscopy \cite{yadav2016observation, das2019observation} and by x-ray coherent diffractive imaging \cite{shao_arxiv_2022}. These vortex structures offer the promise of creating new electronic states of matter, with structured domains with nanometer-scale domain sizes.
Similar to PRIMES, we do not aim to recover the full atomic structure of the material.
Instead, we model the scattering matrix using parameters \revise{(local tilt and polarization)} that represent small perturbations from an \textit{a priori} known average structure of the material, and refine the model to match measured diffraction intensities.
Parameterizing the scattering model allows us to choose a small and physically meaningful set of variables to refine against, and by computing gradients of the model semi-analytically we are able to dramatically accelerate the discovery of the model parameters. 
\revise{Because mistilt is specifically included as a parameter of the model, slight deviation from the zone axis configuration does not couple into errors in the other recovered signals.
By computing the full scattering matrix we are also able to include all beams present in the experiment, enabling the method to work at high-symmetry orientations.}
\revise{By operating on shallow convergence angle nanobeam diffraction patterns, we are able to approximate the diffraction condition as a plane-wave experiment, enabling us to use a smaller \smat that is faster to compute than CBED-based methods such as QCBED, at the cost of reduced sensitivity from losing the rich phase contrast information present in a CBED pattern.}
As a result, our method is sufficiently fast to allow us to perform the parameter matching for each probe position in a four-dimensional scanning transmission electron microscopy (4D-STEM) scan, where a shallow converged electron probe is rastered \revise{in a 500$\times$400 pixel grid} across the sample surface. 
\revise{We find that an advantage of using this dynamical refinement procedure to obtain the local tilt and polarization is that we are able to robustly disentangle these two signals, which often confound one another when measaured using conventional approaches, especially for the multi-layer system we examine. }

\revise{In the Bloch wave description of electron diffraction, which we utlilize in this work, the \smat is computed from the Fourier coefficients of the crystal electrostatic potential, the beam direction, and the sample thickness.
Thus, our approach can capture a wide variety of structural distortions, such as polar distortions, subtle phase changes, chemical (dis)ordering, and other atomic rearrangements that do not distort the locations of the reciprocal lattice points.
The stacked \smat model is further able to incorporate the thickness and tilt of each layer independently.
Properties such as strain that \textit{distort} the (reciprocal) lattice in a continuous manner without substantial change in the Fourier coefficients, or major phase changes that alter the (reciprocal) lattice in an abrupt manner, are not included in this approach as the entries of the \smat cannot be expressed in terms of these parameters.
}

\section{Theory}
\label{sec:theory}

To compute the dynamical diffraction intensities, we utilize the Bloch wave method, which is fully described in DeGraef \cite{de2003introduction}.
In this method, the electron wave is written as a combination of Bloch states, and thus the Schr\"odinger equation for the fast electron wave is cast as an eigenvalue/eigenvector decomposition
\begin{equation}
    \bar{\mathcal{A}} \mathcal{C} 
    = 
    2 k_n \gamma \mathcal{C},
\label{eq:blogheig}
\end{equation}
where the ``structure'' matrix $\bar{\mathcal{A}}$ is determined by the crystal structure and orientation of the sample, $\mathcal{C}$ is a matrix whose column vectors contain the Bloch wave  coefficients, $k_n$ is the normal component of the incident wavevector, and $ \gamma$ relative normal component of each of the Bloch waves. The entries of the structure matrix are given by
\begin{equation}
\bar{\mathcal{A}} = 
\begin{bmatrix}
0 & U_{-\mathbf{g}} & \cdots & U_{-\mathbf{h}} \\
U_{\mathbf{g}} & 2 k_0 s_\mathbf{g} & \cdots & U_{\mathbf{g}-\mathbf{h}} \\
\vdots & \vdots & \ddots & \vdots \\
U_{\mathbf{h}} & U_{\mathbf{h}-\mathbf{g}} & \cdots & 2 k_0 s_\mathbf{h}
\end{bmatrix}
\end{equation}
where $U_{\vec{g}-\vec{h}}$ is the Fourier component of the sample electrostatic potential corresponding to the scattering vector $\vec{g}-\vec{h}$ and $s_{\vec{g}}$ is the excitation error for the beam $\vec{g}$. 
The excitation error is given by
\begin{equation}
    s_{\vec{g}} = 
    \frac{-\vec{g} \cdot \left( 2 \vec{k} + \vec{g} \right)}
    {2 | \vec{k} + \vec{g} | \cos \alpha}
\end{equation}
where $\vec{k}$ is the wavevector of the incident electron beam, and $\alpha$ is the angle between the sample normal and the incident beam direction. 
To include absorption due to thermal diffuse or inelastic scattering, an imaginary component of the sample potential $U_{\vec{g}}'$ is included in the matrix $\bar{\mathcal{A}}$ by adding $\mi U_{\vec{0}}'$ to the diagonal and $\mi U_{\vec{g}-\vec{h}}'$ to the off-diagonal. 
By computing the eigenvalue/eigenvector decomposition of this matrix, we obtain the Bloch wave coefficients $C_{\vec{g}}^{(j)}$ and the normal components $\gamma^{(j)}$, which are used to obtain the scattered wave amplitudes for a given crystal thickness. The electron wave $\vec{\psi}$ at a depth $z$ in the crystal is
\begin{equation}
\vec{\psi}(z) = \mathcal{C} \mathcal{E}(z) \mathcal{C}^{-1} \vec{\psi}_0 = \mathcal{S}(z) \vec{\psi}_0
\label{eq:bloch-thick}
\end{equation}
where $\mathcal{C}$ is the matrix containing the eigenvectors and $\mathcal{E}(z) = \me^{2 \pi \mi \gamma^{(j)}z} \delta_{ij}$ is a diagonal matrix which depends on the thickness and the Bloch wave normal components. 
$\vec{\psi}_0$ is a vector containing the Fourier coefficients of the incident electron wave---in the case of plane wave illumination, it is a vector with the value of 1 at the index corresponding to the incident beam direction and zero elsewhere. 

This transformation \revise{can be compactly} represented by the scattering matrix $\mathcal{S}$ which maps the vector representing the incident electron wave to the scattered wave at depth $z$. 
Writing the equation in this form, we see that the \smat is the exponential of the structure matrix $\mathcal{\vec{A}}$ multiplied by $2 \pi \mi z$.

\subsection{Stacked \smat model}

In the description of the Bloch wave model above, we obtained a single \smat which transformed a plane wave $\vec{\psi}_0$ incident onto a crystal of some thickness $z$ into the scattered wave $\vec{\psi}$. 
If the electron wave were to immediately enter another crystal, we can model this further scattering by simply using the \smat of that layer to transform the previously scattered wave into the final scattered wave \cite{SINGH201832,PhysRevB.89.205409}. 
This operation is equivalent to multiplying the complex \smats together and applying it once to the original wave. 
From a numerical standpoint, it is more convenient to successively apply the \smats to $\vec{\psi}_0$ rather than multiply the \smats first, as matrix-vector multiplications are cheaper than the matrix-matrix products needed to construct the total \smat.

In this work, we will model the trilayer STO:PTO:STO sample using the product of 3 \smats:
\begin{equation}
    \mathcal{S} = 
    \mathcal{S}_{\text{STO}} \;
    \mathcal{S}_{\text{PTO}} \;
    \mathcal{S}_{\text{STO}}
    \label{eqn:stacked-smat}
\end{equation}

For the trilayer sample we consider here, the epitaxial relationship between the layers considerably simplifies the use of the stacked \smat approach, as the layers share the same lattice---this \revise{ensures that all} of the \smats share the same set of beams. 
In a situation where the layers do not share the same lattice, one must ensure that all of the component \smats include all of the beams from all of the layers. 
Further, in such case it is possible for a beam scattered by the top layer to be evanescent in a lower layer, with exponentially decaying intensity \cite{SINGH201832}.

\subsection{Derivatives of the stacked \smat model}

Using the Bloch wave method, we have obtained a scattering matrix which is the exponential of the structure matrix, whose entries are readily obtained from the crystal properties.
The full scattering matrix describing the multilayer  is represented as a product of three such \smats.
In order to match the parameters of the model to our experimental data, we will use an optimization procedure to minimize the error between the model and the experiment.
Unfortunately, each evaluation of the scattered wave using this model requires (for each layer) building a new $\mathcal{A}$-matrix and diagonalizing it, which is computationally expensive and makes numerical optimization inefficient.
Therefore we aim to obtain the derivatives of the scattering matrix with respect to the entries of the structure matrix, so that we can compute the gradient of the error without the large number of function evaluations necessitated by finite differences. 
Najfeld and Havel \cite{NAJFELD1995321} define the directional derivative $D_{\vec{V}}(t,\vec{A})$ of a matrix exponential $e^{t\vec{A}}$ in the direction $\vec{V}$ as
\begin{equation}
    D_{\vec{V}}(t,\vec{A}) 
    \equiv 
    \lim_{h \rightarrow 0} \frac1h 
    \left( 
        e^{t(\vec{A}+h\vec{V})} - e^{t\vec{A}},
    \right)
\end{equation}
We will also use the notation $\frac{d \mathcal{S}}{d \theta}$ to refer to the derivative of a scattering matrix in the direction $\vec{V}_{\theta}$, where $\theta$ is one of the structural perturbation parameters. 
For a matrix which has been spectrally decomposed as $\vec{A} = \vec{U} \vec{\Lambda} \vec{U}^{-1}$, (i.e. its eigendecomposition has been computed, where the columns of $\vec{U}$ contain the eigenvectors and the diagonal of $\vec{\Lambda}$ contains the eigenvalues $\lambda_i$) the directional derivative of its exponential can be computed as \cite{NAJFELD1995321}
\begin{equation}
    D_{\vec{V}} (t, \vec{A}) 
    =
    \vec{U} 
    \left( 
        \left( 
            \vec{U}^{-1}\vec{V}\vec{U} 
        \right) \odot \vec{\Phi} (t) 
    \right) \vec{U}^{- 1}.
    \label{eq:expderiv}
\end{equation}
where $\odot$ is the Hadamard, or elementwise, product. The entries of $\vec{\Phi}(t)$ depend on the eigenvalues of $\vec{A}$:
\begin{equation}
\Phi_{i j} (t) = \left\{\begin{array}{ll}
  (e^{t \lambda_i} - e^{t \lambda_j}) / (\lambda_i - \lambda_j) & \text{if~}
  \lambda_i \neq \lambda_j\\
  t e^{t \lambda_i} & \text{if~} \lambda_i = \lambda_j
\end{array}\right.
\end{equation}

To compute the derivative of the total \smat comprised of an ordered collection of $N$ separate scattering matrices indexed with the
superscript $^{(j)}$ (i.e. $\mathcal{S} = \prod_j \mathcal{S}^{(j)}$), with
respect to a parameter $\theta$ we use the product rule
\begin{equation}
    \frac{d \mathcal{S}}{d \theta} = \sum^N_{i=0} \left[ \prod_{j = 0}^{j = i - 1}
\mathcal{S}^{(j)} \cdot \frac{d \mathcal{S }^{(i)} \mathcal{}}{d \theta} \cdot
\prod^{k = N}_{k = i + 1} \mathcal{S}^{(k)} \right].
\label{eq:productrule}
\end{equation}
Due to our choice of parameters for the model, many of the derivatives $\frac{d \mathcal{S}^{(j)}}{d \theta}$ will be zero. 
For the terms where there are nonzero derivatives, the derivative of the total \smat comprises the scattering up to the layer affected by the parameter $\theta$, the change in scattering within that layer, and the further scattering of the wave by the following layers in the heterostructure. 

In our model of the trilayer sample, there are two relevant classes of structural perturbation that we will attempt to recover: crystal tilt and structural distortion due to polarization. In the following section, we will derive the direction matrices $\vec{V}_{\theta}$ for these types of perturbation.

\subsubsection{Crystal tilt}
Because tilt of the crystal is included solely in the diagonal elements of the structure matrix via the excitation errors $s_{\vec{g}}$, the derivative direction is
\begin{equation}
\left(\vec{V}_{\text{tilt}}\right)_{\vec{g},\vec{h}} = 2 k_0 \frac{d s_{\vec{g}}}{d \vec{k_0}} \delta_{\vec{g} - \vec{h}}
\end{equation}
The derivative of $s_{\vec{g}}$ with respect to the transverse ($x$ and $y$) components of the incident wavevector are approximately
\begin{equation}
\frac{d s_{\vec{g}} }{d k_{\{x,y\}}} = 
- \frac{g_{\{x,y\}}}{| \vec{g} + \vec{k_0} |} + 
\frac{(g_{\{x,y\}} + k_{\{x,y\}}) (\vec{g} \cdot (2
\vec{k} + \vec{g}))} {2 | \vec{g} + \vec{k} |^3}.
\label{eq:tiltgrad}
\end{equation}
We have neglected the $\cos\alpha$ term in the denominator of $s_{\vec{g}}$ in this derivation to greatly simplify the expression at the cost of some small error in the magnitude of $\frac{ds}{dk}.$

\subsubsection{Polarization}

While it is possible under certain conditions to recover the locations of each atom in the unit cell by recovering the full \smat from diffraction data, \cite{PhysRevLett.125.065502,brown2018structure} here we parameterize the model in terms of the polarization directly, and displace the Ti and O sites by interpolating between their positions in the canonical non-polar and polar structures. 
For each atom in the unit cell, we define displacement vectors $\delta \vec{r}_a^{(j)}$ and $\delta \vec{r}_b^{(j)}$ which takes the atom from its site $\vec{r}^{(j)}$ in the non-polar structure to its site in the canonical polar structure, for polarizations in the $a$ and $b$ directions respectively. 
\revise{This fully polar structure for PTO is related to the nonpolar one by displacing the Pb site by 0.0281 lattice units, the $(\frac12, 0, \frac12)$ and $(\frac12, \frac12, 0)$ O sites by -0.0849, and the $(0, \frac12, \frac12)$ O site by -0.1058, while leaving the Ti sites undisturbed, in the case of $a$-axis polarization.}
For intermediate or mixed-direction polarizations, we linearly scale the displacement vectors by the relative polarization, $\rho_a$ and $\rho_b$ for $a$ and $b$ polarizations respectively.
The Fourier coefficients of the crystal potential are thus written as 
\begin{equation}
    U_{\vec{g}} = \frac{1}{\Omega} \sum_j f_e^{(j)} e^{2 \pi i ((\vec{r}^{(j)} + \rho_a \delta \vec{r}_a^{(j)} + \rho_b \delta \vec{r}_b^{(j)})
\cdot \vec{g})}
\end{equation}
where $\Omega$ is the unit cell volume, and the atomic form factors $f_e^{(j)}$ are computed using the absorptive Weickenmeier-Kohl parameterization for isolated neutral atoms \cite{weickenmeier1991computation}.
The derivative of the structure factor with respect to the relative $a$-axis polarization parameter, $\rho_a$, is then
\begin{equation}
    \frac{d U_{\vec{g}}}{d \rho_a} 
    = 
    \frac{1}{\Omega} \sum_j 2 \pi i f_{e }^{(j)} (\vec{g} \cdot \delta
    \vec{r}_a) e^{2 \pi i ((\vec{r}^{(j)} + \rho_a \delta \vec{r}_a^{(j)} + \rho_b \delta \vec{r}_b^{(j)}) \cdot \vec{g})}
    \label{eq:polargrad}
\end{equation}
and a similar expression arises for $\frac{d U_{\vec{g}}}{d\rho_b}$. 
The derivative direction matrix for polarization is simply filled with the derivatives of the Fourier coefficients, e.g.
\begin{equation}
    \left( \vec{V}_{\rho_{\{a,b\}}} \right)_{\vec{g},\vec{h}} = \frac{d U_{\vec{g}-\vec{h}}}{d \rho_{\{a,b\}}}
\end{equation}
Note that we do not take into account expansion of the unit cell, so the mean inner potential term $U_{\vec{0}}$ does not change, and thus polarization only affects the off-diagonal elements of the structure matrix. 

\subsection{Numerical Optimization}
To fit the stacked \smat model to the experimental measurements, we implemented a version of the alternating direction method of multipliers \cite{boyd2011distributed}. 
At each iteration of the optimization algorithm, we first update the model parameters at each probe position using our previously derived gradients and taking a step along the direction of steepest descent. 
We then perform regularization of the fitted model parameters to ensure convergence to a physically sensible solution and enforce smoothness.

\subsubsection{Gradient Descent}
The loss function $\mathcal{L}$ is the sum squared difference between the simulated diffraction intensities from the model and the experimental intensities for each Bragg beam recorded
\begin{equation}
    \mathcal{L} = \sum_{\vec{g}} \left( I_{\text{exp}}(\vec{g}) - \mu | \mathcal{S} \vec{\psi}_0(\vec{g}) |^2 - \nu \right)^2
\end{equation}
The modeled intensities have both an additive intensity offset $\nu$ and multiplicative scaling $\mu$, which we found necessary in order to compensate for background noise and intensity variation in the experimental data. 
Note that the \smat calculations are performed using a different, and larger, set of Bragg beams, in order to include scattering into the higher order beams (not recorded on the detector) in the forward model. 
\revise{Our experimental diffraction patterns measure the intensities of 69 Bragg beams, with a maximum scattering vector of 1.1~\AA$^{-1}$.
The \smat calculations include 109 beams, with a maximum scattering angle of 1.5~\AA$^{-1}$.}
Only the Bragg beams present in the experiment contribute to the loss function.
The derivatives of the loss function with respect to the intensity scale parameters are given as
\begin{align}
    \frac{\partial \mathcal{L}}{\partial \mu} &= \sum_{\vec{g}}
        -2 | \mathcal{S} \vec{\psi}_0 |^2 \left( I_{\text{exp}} - \mu | \mathcal{S} \vec{\psi}_0 |^2 - \nu \right) \\
    \frac{\partial \mathcal{L}}{\partial \nu} &= \sum_{\vec{g}}
        -2 \left( I_{\text{exp}} - \mu | \mathcal{S} \vec{\psi}_0 |^2 - \nu \right)
\end{align}
(where we have dropped the dependence on $\vec{g}$ from the notation for compactness).
The derivatives with respect to the structure perturbations involve the derivatives of the \smat, and so are much more complicated expressions. For a generic parameter $\theta$ that enters into the \smats, the derivative of the loss function is
\begin{equation}
    \frac{\partial \mathcal{L}}{\partial \theta} = 
    -4 \mu \, \mathfrak{Re}\left[ 
    \vec{\psi}^* \frac{d \vec{\mathcal{S}}}{d \theta} \vec{\psi}_0 
    \right] \cdot \left[
    I_{\text{exp}} - \mu | \mathcal{S} \vec{\psi}_0 |^2 - \nu
    \right]
\end{equation}
The gradients with respect to the tilt and polarization variables are obtained using the derivative directions in Eqs.~\ref{eq:tiltgrad} and \ref{eq:polargrad}, the product rule in Eq.~\ref{eq:productrule}, and the \smat derivative method in Eq.~\ref{eq:expderiv}.
At each step of the optimization procedure, we update the parameters by taking a step along the negative gradient direction of this loss function.
\revise{In some previous works, such as those based on quantitative CBED \cite{jm1998quantitative}, optimization would begin by first refining the geometrical factors in the model such as tilt and thickness before proceeding to refine structural distortions such as polarization. In this work, we instead begin updating all parameters from the start of iteration but use regularization in order to rapidly and robustly converge on a solution.  }

\subsubsection{Regularization}

In order to obtain physically sensible solutions to the optimization problem we found it necessary to apply several regularizers.
Before performing the optimization, we de-noise the integrated disk intensities using principal component analysis, retaining the first 16 components.

At each iteration step, we apply further regularization. 
First, the estimated parameters are smoothed across the real-space dimensions of the scan using a Gaussian kernel. Since the intensity scale and offset and the tilts are expected to vary slowly across the field of view, we used a kernel size of 50 nm for the intensity parameters and 25 nm for the tilts. The polarization is expected to vary more rapidly, so we used a kernel size of 2 nm. Note that the experiment used a probe step size of 1 nm, giving equivalent values for the size of each kernel in terms of the number of probe positions.

In addition, we also clip the fitted parameters to be within set bounds, so that outliers do not excessively propagate error to their neighbors via the smoothing kernel. 
We note that we do not apply any explicit high-pass filtering to the fitted polarization values (on the contrary, they are Gaussian filtered, albeit with a very small kernel size). 
However, the strong smoothing regularization applied to the tilt and intensity signals can have the side effect of forcing all of the high frequency variation into the polarization channel. 

\subsection{Atomic Form Factors}
Wu et al \cite{WU2020113095} showed that 4D-STEM may be sensitive to the charge transfer between sites in ionic materials, using strontium titanate as a model system, which would imply that the independent atom model for the crystal potential may not be valid for our computations.
To test this possibility we used the GPAW density functional theory package \cite{PhysRevB.71.035109} to simulate the charge transfer between species, and then used abTEM \cite{10.12688/openreseurope.13015.2} to perform diffraction simulations for the simulated charge densities that match our experimental conditions. 
From these simulations we observed that the maximum deviation in the diffracted intensities between the DFT and IAM potentials was approximately 0.1\% of the probe intensity, validating the use of the IAM model for our computations.

\section{Methods}

\subsection{Heterostructure Growth}

We synthesized a trilayer structure consisting of 16 unit cells of SrTiO$_3$, 16 unit cells of PbTiO$_3$, and 16 unit cells of SrTiO$_3$, on top of a SrRuO$_3$ buffer layer on a single crystal DyScO$_3$ substrate, 
\revise{giving a total heterostructure thickness of  $\approx 19$~nm.}
The layers were grown at 610$^\circ$C in a 100 mTorr oxygen atmosphere, using reflection high-energy electron diffraction (RHEED)-assisted pulsed laser deposition (PLD) with a KrF laser.  
\revise{The trilayer structure was prepared for TEM analysis by mechanical polishing followed by ion milling.
Within the imaged area the substrate has been completely removed, leaving only the heterostructure. }

\subsection{4D-STEM Experiments}

We performed 4D-STEM measurements on the TEAM I microscope, an aberration-corrected Thermo Fisher Scientific Titan operated at 300 kV with a probe current of 100 pA. 
We used a STEM probe semiangle of 2~mrad, and a STEM probe step size of 1~nm. 
We recorded \revise{zero-loss filtered} diffraction patterns using a Gatan K3 direct electron detector located beyond a Gatan Continuum energy filter. 
We operated the K3 detector in electron counting mode using a binning of 4x4 pixels, a camera length of 1.05~m, and an exposure time of 47~ms. 
We analyzed the 4D-STEM experiments using custom Python and Matlab code. 
The diffraction pattern simulations and Bloch wave calculations and optimizations have been implemented as part of the py4DSTEM analysis toolkit \cite{savitzky2021py4dstem, ophus2022automated, zeltmann20224dstem}. 
\revise{Initial processing of the data was performed by selecting the 69 Bragg disks visible on the detector and integrating their intensities. 
All further analysis was conducted using these integrated disk intensities rather than the pixelwise intensity of the recorded patterns.
As a result, the analysis is not expected to be sensitive to distortion by the post-specimen optics or the point-spread function of the detector.}

\section{Results \& Discussion}

\subsection{Tilt/Polarization Confounding}
\begin{figure}[tbp]
    \centering
    \centerline{\includegraphics[width=1.2\textwidth]{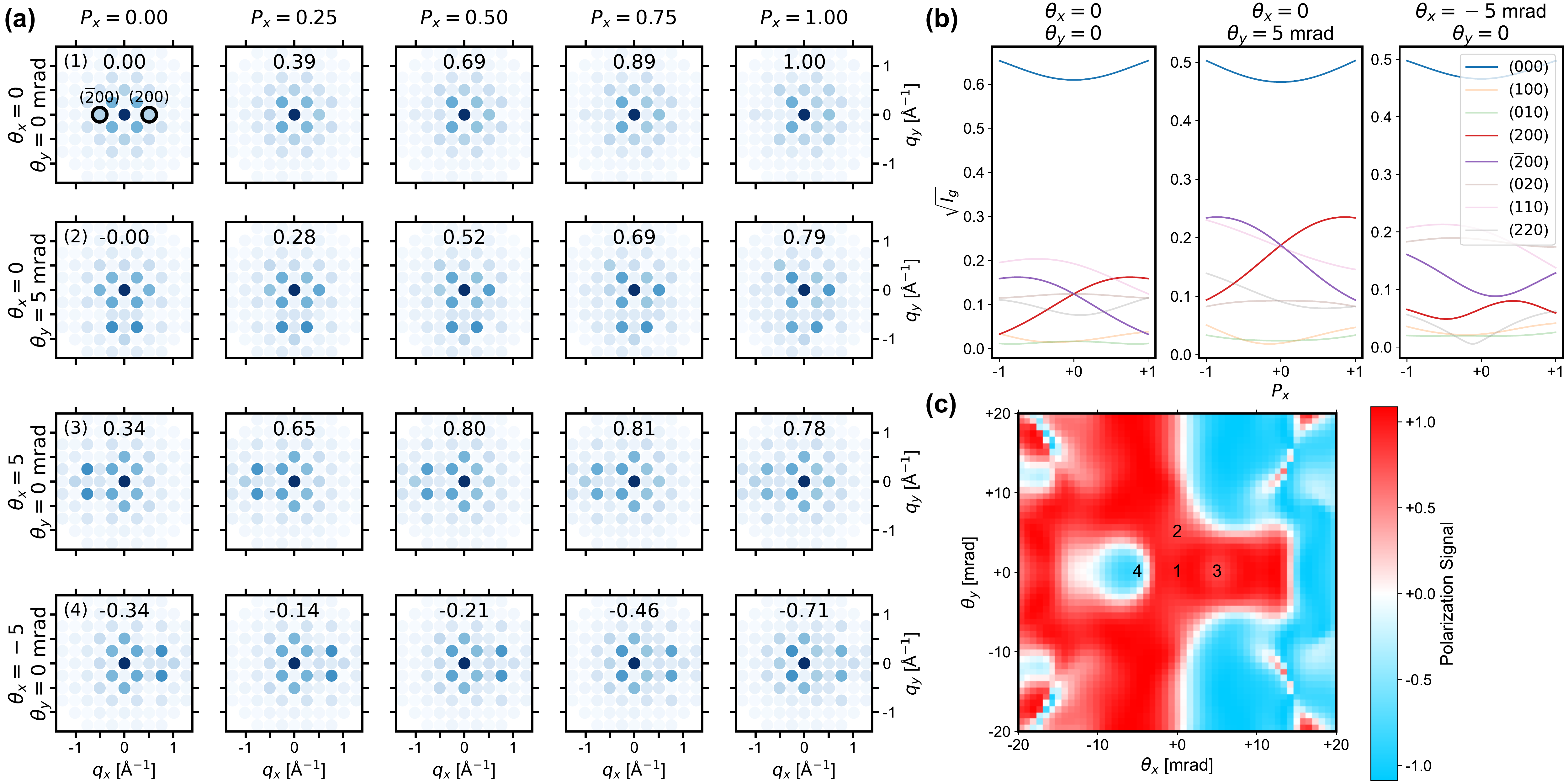}}
    \caption{Confounding of tilt and polarization signals for a STO:PTO:STO trilayer. (a) Sequences of diffraction patterns simulated at relative $x$-direction polarization varying from 0--1 for various mistilts from a perfect $[001]$ orientation. The two diffraction disks conventionally used to measure polarization, $(200)$ and $(\overline{2}00)$ are highlighted, and the inset text indicates the relative polarization measured as $\frac{I_{200}-I_{\overline{2}00}}{I_{200}+I_{\overline{2}00}}$. (b) Line traces of selected diffraction intensities for different mistilts, with the $(200)$ and $(\overline{2}00)$ reflections highlighted. (c) Map of apparent polarization signal for a trilayer with $P_x=1$ as measured from the asymmetry of the $(200)$ and $(\overline{2}00)$ disks for different mistilts. Contrast reversals in the polarization signal occur with at little as 5~mrad mistilt. The overlaid numbers indicate the tilt values corresponding to the rows of (a).}
    \label{fig:til_polar_coupling}
\end{figure}


Figure~\ref{fig:til_polar_coupling} shows how local mistilts of the sample from the perfect zone axis orientation can confound the measurement of local polarization when using a conventional metric based on Friedel pair asymmetry. 
The top row of Fig.~\ref{fig:til_polar_coupling}a shows a sequence of simulated diffraction patterns for the STO/PTO trilayer sample at varying polarization in the $x$-direction. 
The inset numbers indicate the polarization signal measured using the anomalous contrast of the $(200)$ Friedel pair \cite{deb2020imaging}, computed as $(I_{200}-I_{\overline{2}00})/(I_{200}+I_{\overline{2}00})$ and normalized to the $\theta_x=\theta_y=0$, $P_x=1$ value. 
The signal is monotonic with increasing polarization and approximately linear, indicating that in the ideal case the symmetry breaking of this pair of diffraction disks is a good measurement of the local polarization. 
As shown in the left panel of Fig.~\ref{fig:til_polar_coupling}b, the intensities of these disks branch as a function of polarization. 
When tilting the incident beam towards the positive $y$-axis, as in the second row of Fig.~\ref{fig:til_polar_coupling}a, the symmetry of $(200)$ pair of diffraction disks is not broken, but the different excitation of these beams (shown in the center panel of Fig.~\ref{fig:til_polar_coupling}b) causes the signal to be suppressed by approximately 20\%. 
However, when the beam is tilted by 5~mrad ($\approx0.3^{\circ}$) towards the $x$-axis (as shown in the right panel of Fig.~\ref{fig:til_polar_coupling}b), the effect of the tilting is to break the symmetry of the $(200)$ disks. 
This slight mistilt in either direction along the $x$-axis completely destroys the polarization measurement.
In the third row of Fig.~\ref{fig:til_polar_coupling}a, where the beam is tilted towards the positive $x$-axis, there is an apparent nonzero polarization even when the material is not polarized.
In the fourth row of Fig.~\ref{fig:til_polar_coupling}a (and the right panel of Fig.~\ref{fig:til_polar_coupling}b) a tilt towards the negative $x$-axis causes an inversion of the polarization signal when $P_x>0$, non-monotic behavior when $P_x<0$, as well as an apparent nonzero polarization even when the material is not polarized.
These effects are plotted as a function of $x$ and $y$ tilts in Fig.~\ref{fig:til_polar_coupling}c.

\subsection{Gradients of the Diffracted Intensities}

\begin{figure}[tbp]
    \centering
    \centerline{
    \includegraphics[width=1.2\textwidth,valign=c]{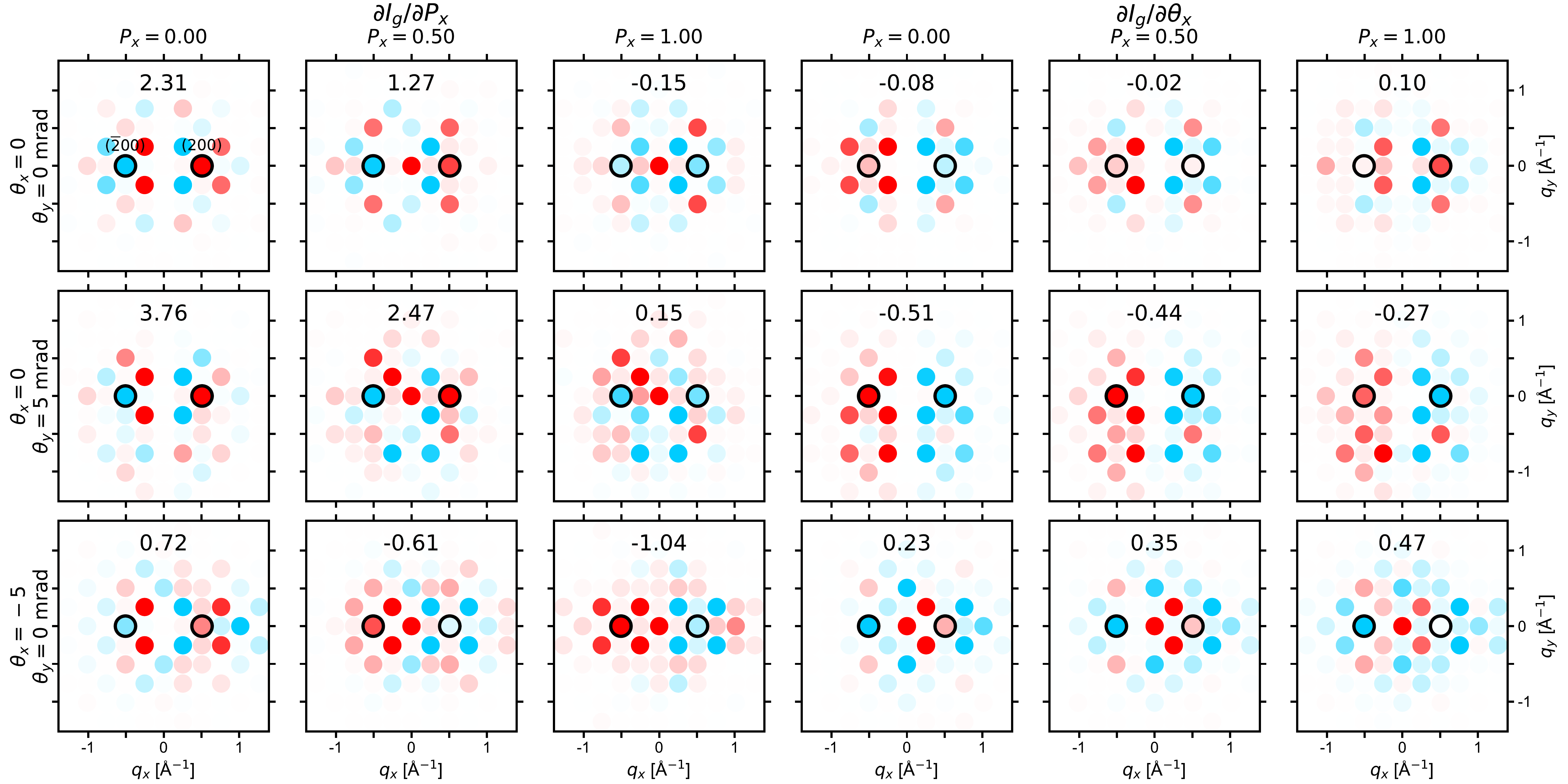}
    \includegraphics[width=0.06\textwidth,valign=c]{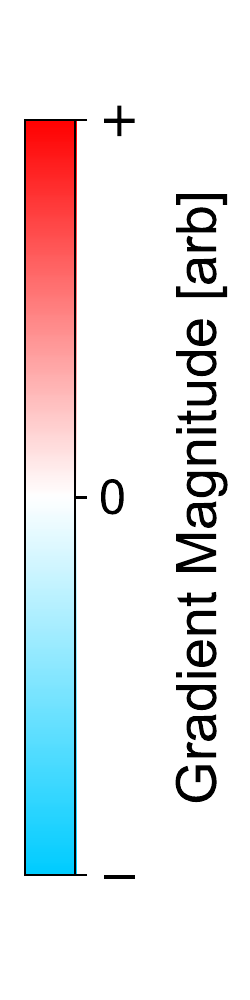}
    }
    \caption{Derivatives of the diffracted intensities with respect to $x$-polarization $P_x$ and tilt about the $x$-axis $\theta_x$. \revise{The inset numbers show the value of $\frac{\partial I_{200}}{\partial \vartheta} - \frac{\partial I_{\overline{2}00}}{\partial \vartheta}$ in arbitrary units, where $\vartheta$ represents polarization for the left 3 columns, and tilt for the right 3 columns.}}
    \label{fig:jacobians}
\end{figure}

Computations of the derivatives of the diffraction disk intensities with respect to  $x$-direction polarization and tilt are shown in Figure~\ref{fig:jacobians}, evaluated over a range of tilts and polarizations.
The inset numbers indicate the difference between the derivative of the $(200)$ Friedel pair, which demonstrates the sensitivity of the signal derived from the anomalous contrast of those reflections to the chosen parameter.
The difference in overall magnitude between the polarization and tilt derivatives is affected by the choice of units for the parameters; in the figure they have been scaled to be visually uniform, and in the optimization procedure the problem is rescaled to promote uniform convergence along all the parameter directions. 
In the on-zone, unpolarized case ($P=0$, $\theta=0$) in the top left, the $(200)$ anomalous contrast signal will not distinguish between polarization of the crystal and tilt, as both cause the same anomalous contrast. 
However, other reflections respond in different ways to polarization and tilt. 
Thus, when considering all of the diffracted beams the gradient directions for tilt and polarization are approximately 62$^\circ$ separated.
Since they are not orthogonal, an iterative optimization will be needed in order to solve the polarization and tilt.
In the case of nonzero polarization and tilt, the gradients tend to become more parallel.
In particular the gradients with respect to $y$-direction polarization and tilt, which are fully orthogonal to the $x$-direction parameter gradients at $P=0$, $\theta=0$, will become partially coupled to the other direction when the crystal is tilted or polarized. 

\subsection{Experimental Results}

\begin{figure}[tbp]
    \centering
    \centerline{\includegraphics[width=1.0\textwidth]{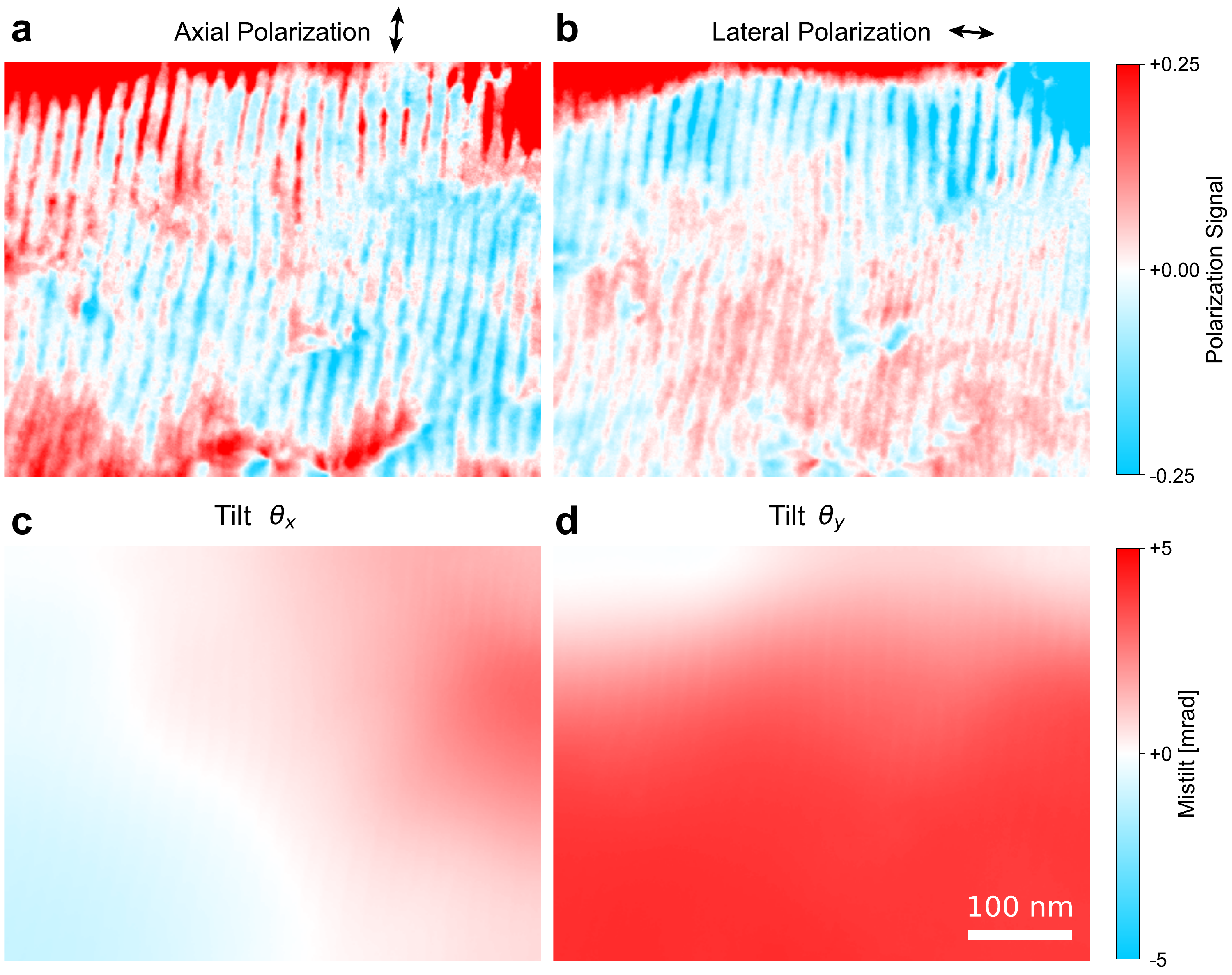}}
    \caption{\revise{(a),(b) Polarization and (c),(d) tilt of the STO:PTO:STO sample recovered from experimental measurements using the optimization procedure.} Approximate polarization directions are labeled above.}
    \label{fig:experiment}
\end{figure}

The fitted polarization and tilts of the STO:PTO:STO multilayer sample are plotted in Fig.~\ref{fig:experiment} over the full field of view. Immediately, we can see several domains in the polarization maps in Figs.~\ref{fig:experiment}a and b where the regular periodic structures show the tubular vortex structures. The ``axial'' polarization represents the PTO polarization along the directions parallel to the vortex cores, while the ``lateral'' polarization is perpendicular. Inside each vortex, the projected polarization is relatively constant in a given domain. Various domain boundaries are also visible, where the polarization abruptly changes sign, in either the axial or lateral directions, or both. These domain structures and domain walls are in good agreement with previous observations of STO/PTO multilayer samples \cite{yadav2016observation, das2019observation, susarla2022emergence}. The tilt maps shown in Figs.~\ref{fig:experiment}c and d show significant rotation from the ideal zone axis, especially in the $y$-axis direction. These maps demonstrate the need to include tilt in the modeling of the diffraction signals. The smoothness of the estimated tilt is due in part to the strong regularization applied during the reconstruction.

The polarization maps contain many complex domain and domain wall structures. We expect that the vortex cores will have alternating polarization signs in the axial direction. This alternating structure is visible in all domains in Fig.~\ref{fig:experiment}a, though interestingly we also observe a negative offset from zero mean axial polarization in the largest domain spanning the grain in the bottom half of the map. The grains at the top and bottom edges also show a significant positive offset from zero mean. These observations suggest that there may be a net axial polarization in many of the domains, which can't be directly observed from qualitative estimates of the polarization which have been high-pass filtered \cite{susarla2022emergence}. 

By contrast, in the lateral direction we expect oscillations in the polarization, but that each domain will have a larger net positive or negative polarization. This is because phase field predictions of the polarization structure of the PTO vortex phase predict that that every other vortex will be displaced towards one of the STO/PTO interfaces, while the remaining vortices will be displaced towards the other PTO/STO interface \cite{susarla2022emergence}. This in turn causes a net polarization flow to one of the lateral directions. Susarla et al.~ provide more phase field modeling and predicted vortex domain structures \cite{susarla2022emergence}. These net polarization features are indeed observed in Fig.~\ref{fig:experiment}b, \revise{and the overall magnitude of the measured polarization is similar to those observed previously}. Domains in the top third of the map and bottom left show significant polarization towards the negative direction, while the domains in the bottom two thirds show significant polarization in the positive direction. Various small domains are interspersed into the larger domains, but each shows a non-zero mean polarization. Overall, these observations provide a significant step forward in accurate modeling of the intensity of Bragg peaks when the beam undergoes significant multiple scattering and the sample has a large mistilt from the ideal zone axis.

\begin{figure}[tbp]
    \centering
    \centerline{\includegraphics[width=0.6\textwidth]{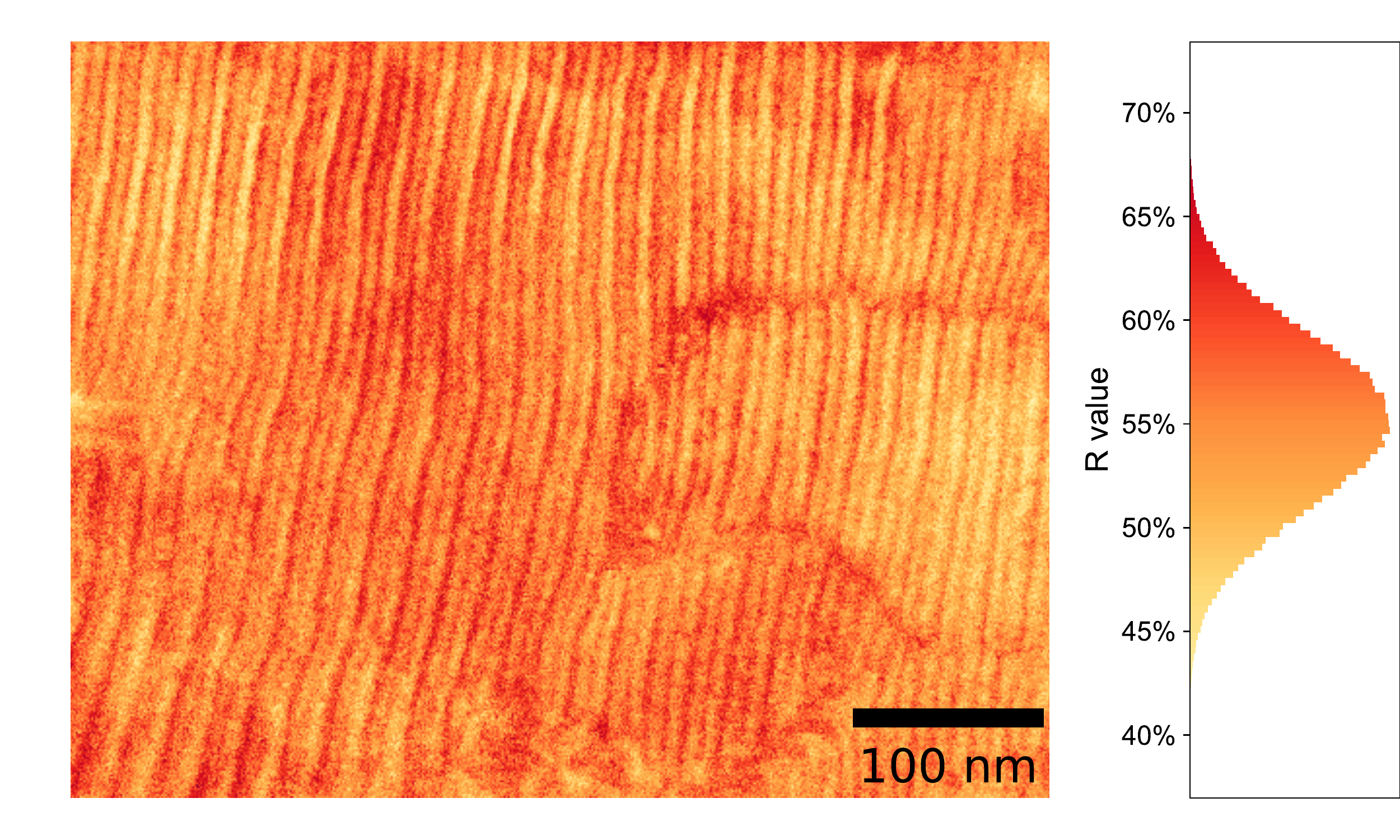}}
    \caption{\revise{$R$ value of the fitted model at each scan position.  }}
    \label{fig:error}
\end{figure}

\revise{
The goodness of fit of the model is plotted for each scan position in Fig.~\ref{fig:error}, using the residue, or $R$-value, defined as
\begin{equation}
R = \frac{\sum_{\vec{g}} \left| I_{\text{exp}}(\vec{g}) - \mu | \mathcal{S} \vec{\psi}_0(\vec{g})|^2 \right| }{\sum_{\vec{g}} I_{\text{exp}}(\vec{g}) }
\end{equation}
Overall the $R$ values are relatively high, with an average of about 55\%, though they are consistent across the field of view. 
The overall high value of this residue may point to a systematic mismatch between the experiment and the model.
Surface contamination, a small amount of remaining buffer layer from the synthesis procedure, or the presence of an amorphous damage layer could all contribute to another source of diffracted intensity that cannot be accounted for by the model.
Incorrect estimates of the Debye-Waller factors also change the distribution of intensity between the different orders of diffraction, leading to high residue values but not preventing the local parameter variations from being recovered well.
The application of strong regularization on the recovered parameters is also likely to lead to larger residue values.
A better fit could be achieved by tuning global parameters such as the Debye-Waller factors, regularization strength, or weighting of intensities in the loss function, though this would add considerably to the computational complexity.
Recent advances in interpretable machine learning may offer an efficient means for tuning these hyperparameters, in particular the methods known as ``algorithm unrolling,'' which provides a connection between iterative algorithms like the one we use here, and deep learning frameworks \cite{monga2021algorithm}. 
In addition, there is finer variation in the residue which corresponds to the fluctuations in polarization. 
As we discuss in the following section, the structure of the polarization in this sample is more complex than the model we have used can capture, as it varies along the beam direction as well as in-plane. 
}

\subsection{Future Directions}

In this work we have shown a stacked \smat model for scattering through a trilayer heterostructure parameterized with a single homogeneous polarization direction within the PTO layer. 
However, the samples we investigated are known to have polarization that varies in a complex manner through the thickness. 
A natural extension of Equation~\ref{eqn:stacked-smat} to account for this is to model the PTO layer with a product of $N_L$ \smats with distinct polarization:
\begin{equation}
    \mathcal{S}_{\text{PTO}} = \prod_i^{N_L} \widehat{\mathcal{S}}_{\text{PTO}}
(\rho_{a,i}, \rho_{b,i} ; t = T / N_L)
\end{equation}
where $\rho_{a,i}$ and $\rho_{b,i}$ are the relative polarization of the $i$-th layer in the $a$ and $b$ directions, and $T$ is the total thickness. 
This modification to the model allows us to more accurately reproduce the physics of the scattering, at the cost of adding substantially more optimization variables. 
This added complexity can be mitigated somewhat by applying constraints to the variation in polarization with thickness. 
For example, if the polarization is constrained to vary linearly then the polarization variables at the $i$-th layer are expressed in terms of just two optimization variables, $\rho_{\text{top}}$ and $\rho_{\text{bottom}}$ regardless of the number of layers modeled
\begin{equation}
    \rho_i = \rho_{\text{top}} + \frac{i}{N_L - 1} (\rho_{\text{bottom}} -
\rho_{\text{top}})
\end{equation}
Pennington and Koch \cite{PENNINGTON201542} have used a similar stacked model to solve for polarization changes along the beam direction, but their approach relies on having a ``composite'' CBED measurement that spans a large range of incident beam directions. 
This approach is both experimentally more challenging and several orders of magnitude more computationally intensive, but the inclusion of many beam directions may be necessary in order to obtain 3D information. 
It may also be possible to use multibeam electron diffraction \cite{hong2021multibeam} to obtain several nanobeam diffraction patterns with large angular separation simultaneously.

\section{Conclusions}

In this work, we have constructed a model of the electron multiple scattering through a complex multi-layer sample, parameterized over the physically relevant variables, and utilized an optimization procedure to fit the model to experimental data. Simple models for measuring the polarization of materials from nanobeam electron diffraction patterns, using symmetry breaking of pairs of diffracted disks, break down in the presence of even small tilts of the crystal, causing contrast changes and reversals. Using a stacked \smat approach, we are able to use all of the scattered beams to determine polarization and tilt simultaneously. In order to make the problem computationally feasible for a large area scan, we derived the analytic gradients of the diffraction intensities and used them to perform regularized gradient descent.

\section{Acknowledgements}

SEZ and CO thank Alireza Sadri for reference to the literature on matrix exponential derivatives, Scott Findlay for suggesting the gradient approach and proofreading the manuscript, Toma Susi for useful discussions regarding the use of density functional theory in diffraction simulations, and Christoph Koch for discussions on the stacked Bloch wave approach. SEZ was supported by the National Science Foundation under STROBE Grant No. DMR 1548924. HGB acknowledges the support of a University of Melbourne early career researcher award. CO acknowledges support of a US Department of Energy Early Career Research Award. Work at the Molecular Foundry was supported by the Office of Science, Office of Basic Energy Sciences, of the U.S. Department of Energy under Contract No. DE-AC02-05CH11231. 

\bibliographystyle{model1-num-names}
\bibliography{4DSTEMrefs.bib}

\begin{thebibliography}{45}
\expandafter\ifx\csname natexlab\endcsname\relax\def\natexlab#1{#1}\fi
\providecommand{\bibinfo}[2]{#2}
\ifx\xfnm\relax \def\xfnm[#1]{\unskip,\space#1}\fi
\bibitem[{Yang et~al.(2017)Yang, MacLaren, Jones, Martinez, Simson, Huth, Ryll,
  Soltau, Sagawa, Kondo et~al.}]{yang2017electron}
\bibinfo{author}{H.~Yang}, \bibinfo{author}{I.~MacLaren},
  \bibinfo{author}{L.~Jones}, \bibinfo{author}{G.~T. Martinez},
  \bibinfo{author}{M.~Simson}, \bibinfo{author}{M.~Huth},
  \bibinfo{author}{H.~Ryll}, \bibinfo{author}{H.~Soltau},
  \bibinfo{author}{R.~Sagawa}, \bibinfo{author}{Y.~Kondo}, et~al.,
\newblock \bibinfo{title}{Electron ptychographic phase imaging of light
  elements in crystalline materials using wigner distribution deconvolution},
\newblock \bibinfo{journal}{Ultramicroscopy} \bibinfo{volume}{180}
  (\bibinfo{year}{2017}) \bibinfo{pages}{173--179}.
\bibitem[{Seki et~al.(2022)Seki, Khare, Murakami, Toyama,
  S{\'a}nchez-Santolino, Sasaki, Findlay, Petersen, Ikuhara, and
  Shibata}]{seki2022linear}
\bibinfo{author}{T.~Seki}, \bibinfo{author}{K.~Khare}, \bibinfo{author}{Y.~O.
  Murakami}, \bibinfo{author}{S.~Toyama},
  \bibinfo{author}{G.~S{\'a}nchez-Santolino}, \bibinfo{author}{H.~Sasaki},
  \bibinfo{author}{S.~D. Findlay}, \bibinfo{author}{T.~C. Petersen},
  \bibinfo{author}{Y.~Ikuhara}, \bibinfo{author}{N.~Shibata},
\newblock \bibinfo{title}{Linear imaging theory for differential phase contrast
  and other phase imaging modes in scanning transmission electron microscopy},
\newblock \bibinfo{journal}{Ultramicroscopy} \bibinfo{volume}{240}
  (\bibinfo{year}{2022}) \bibinfo{pages}{113580}.
\bibitem[{Mahr et~al.(2015)Mahr, M{\"u}ller-Caspary, Grieb, Schowalter,
  Mehrtens, Krause, Zillmann, and Rosenauer}]{mahr2015theoretical}
\bibinfo{author}{C.~Mahr}, \bibinfo{author}{K.~M{\"u}ller-Caspary},
  \bibinfo{author}{T.~Grieb}, \bibinfo{author}{M.~Schowalter},
  \bibinfo{author}{T.~Mehrtens}, \bibinfo{author}{F.~F. Krause},
  \bibinfo{author}{D.~Zillmann}, \bibinfo{author}{A.~Rosenauer},
\newblock \bibinfo{title}{Theoretical study of precision and accuracy of strain
  analysis by nano-beam electron diffraction},
\newblock \bibinfo{journal}{Ultramicroscopy} \bibinfo{volume}{158}
  (\bibinfo{year}{2015}) \bibinfo{pages}{38--48}.
\bibitem[{Cooper et~al.(2016)Cooper, Denneulin, Bernier, B{\'e}ch{\'e}, and
  Rouvi{\`e}re}]{cooper2016strain}
\bibinfo{author}{D.~Cooper}, \bibinfo{author}{T.~Denneulin},
  \bibinfo{author}{N.~Bernier}, \bibinfo{author}{A.~B{\'e}ch{\'e}},
  \bibinfo{author}{J.-L. Rouvi{\`e}re},
\newblock \bibinfo{title}{Strain mapping of semiconductor specimens with
  nm-scale resolution in a transmission electron microscope},
\newblock \bibinfo{journal}{Micron} \bibinfo{volume}{80} (\bibinfo{year}{2016})
  \bibinfo{pages}{145--165}.
\bibitem[{Pekin et~al.(2017)Pekin, Gammer, Ciston, Minor, and
  Ophus}]{pekin2017strain}
\bibinfo{author}{T.~C. Pekin}, \bibinfo{author}{C.~Gammer},
  \bibinfo{author}{J.~Ciston}, \bibinfo{author}{A.~M. Minor},
  \bibinfo{author}{C.~Ophus},
\newblock \bibinfo{title}{Optimizing disk registration algorithms for nanobeam
  electron diffraction strain mapping},
\newblock \bibinfo{journal}{Ultramicroscopy} \bibinfo{volume}{176}
  (\bibinfo{year}{2017}) \bibinfo{pages}{170--176}.
\bibitem[{Zeltmann et~al.(2020)Zeltmann, M{\"u}ller, Bustillo, Savitzky,
  Hughes, Minor, and Ophus}]{zeltmann2020patterned}
\bibinfo{author}{S.~E. Zeltmann}, \bibinfo{author}{A.~M{\"u}ller},
  \bibinfo{author}{K.~C. Bustillo}, \bibinfo{author}{B.~Savitzky},
  \bibinfo{author}{L.~Hughes}, \bibinfo{author}{A.~M. Minor},
  \bibinfo{author}{C.~Ophus},
\newblock \bibinfo{title}{Patterned probes for high precision {4D-STEM} {Bragg}
  measurements},
\newblock \bibinfo{journal}{Ultramicroscopy} \bibinfo{volume}{209}
  (\bibinfo{year}{2020}) \bibinfo{pages}{112890}.
\bibitem[{Chen et~al.(2021)Chen, Jiang, Shao, Holtz, Odstr{\v{c}}il,
  Guizar-Sicairos, Hanke, Ganschow, Schlom, and Muller}]{chen2021electron}
\bibinfo{author}{Z.~Chen}, \bibinfo{author}{Y.~Jiang}, \bibinfo{author}{Y.-T.
  Shao}, \bibinfo{author}{M.~E. Holtz}, \bibinfo{author}{M.~Odstr{\v{c}}il},
  \bibinfo{author}{M.~Guizar-Sicairos}, \bibinfo{author}{I.~Hanke},
  \bibinfo{author}{S.~Ganschow}, \bibinfo{author}{D.~G. Schlom},
  \bibinfo{author}{D.~A. Muller},
\newblock \bibinfo{title}{Electron ptychography achieves atomic-resolution
  limits set by lattice vibrations},
\newblock \bibinfo{journal}{Science} \bibinfo{volume}{372}
  (\bibinfo{year}{2021}) \bibinfo{pages}{826--831}.
\bibitem[{Donatelli and Spence(2020)}]{PhysRevLett.125.065502}
\bibinfo{author}{J.~J. Donatelli}, \bibinfo{author}{J.~C.~H. Spence},
\newblock \bibinfo{title}{Inversion of many-beam {Bragg} intensities for
  phasing by iterated projections: Removal of multiple scattering artifacts
  from diffraction data},
\newblock \bibinfo{journal}{Phys. Rev. Lett.} \bibinfo{volume}{125}
  (\bibinfo{year}{2020}) \bibinfo{pages}{065502}.
\bibitem[{Spence et~al.(1994)Spence, Zuo, O'Keeffe, Marthinsen, and
  Hoier}]{spence1994minimum}
\bibinfo{author}{J.~C. Spence}, \bibinfo{author}{J.~Zuo},
  \bibinfo{author}{M.~O'Keeffe}, \bibinfo{author}{K.~Marthinsen},
  \bibinfo{author}{R.~Hoier},
\newblock \bibinfo{title}{On the minimum number of beams needed to distinguish
  enantiomorphs in x-ray and electron diffraction},
\newblock \bibinfo{journal}{Acta Crystallographica Section A: Foundations of
  Crystallography} \bibinfo{volume}{50} (\bibinfo{year}{1994})
  \bibinfo{pages}{647--650}.
\bibitem[{Spence(1998)}]{spence1998direct}
\bibinfo{author}{J.~Spence},
\newblock \bibinfo{title}{Direct inversion of dynamical electron diffraction
  patterns to structure factors},
\newblock \bibinfo{journal}{Acta Crystallographica Section A}
  \bibinfo{volume}{54} (\bibinfo{year}{1998}) \bibinfo{pages}{7--18}.
\bibitem[{Weierstall et~al.(2002)Weierstall, Chen, Spence, Howells, Isaacson,
  and Panepucci}]{weierstall2002image}
\bibinfo{author}{U.~Weierstall}, \bibinfo{author}{Q.~Chen},
  \bibinfo{author}{J.~Spence}, \bibinfo{author}{M.~Howells},
  \bibinfo{author}{M.~Isaacson}, \bibinfo{author}{R.~Panepucci},
\newblock \bibinfo{title}{Image reconstruction from electron and x-ray
  diffraction patterns using iterative algorithms: experiment and simulation},
\newblock \bibinfo{journal}{Ultramicroscopy} \bibinfo{volume}{90}
  (\bibinfo{year}{2002}) \bibinfo{pages}{171--195}.
\bibitem[{M{\"u}ller et~al.(2012)M{\"u}ller, Rosenauer, Schowalter, Zweck,
  Fritz, and Volz}]{muller2012strain}
\bibinfo{author}{K.~M{\"u}ller}, \bibinfo{author}{A.~Rosenauer},
  \bibinfo{author}{M.~Schowalter}, \bibinfo{author}{J.~Zweck},
  \bibinfo{author}{R.~Fritz}, \bibinfo{author}{K.~Volz},
\newblock \bibinfo{title}{Strain measurement in semiconductor heterostructures
  by scanning transmission electron microscopy},
\newblock \bibinfo{journal}{Microscopy and Microanalysis} \bibinfo{volume}{18}
  (\bibinfo{year}{2012}) \bibinfo{pages}{995--1009}.
\bibitem[{Clark et~al.(2018)Clark, Brown, Paganin, Morgan, Matsumoto, Shibata,
  Petersen, and Findlay}]{clark2018probing}
\bibinfo{author}{L.~Clark}, \bibinfo{author}{H.~Brown},
  \bibinfo{author}{D.~Paganin}, \bibinfo{author}{M.~Morgan},
  \bibinfo{author}{T.~Matsumoto}, \bibinfo{author}{N.~Shibata},
  \bibinfo{author}{T.~Petersen}, \bibinfo{author}{S.~Findlay},
\newblock \bibinfo{title}{Probing the limits of the rigid-intensity-shift model
  in differential-phase-contrast scanning transmission electron microscopy},
\newblock \bibinfo{journal}{Physical Review A} \bibinfo{volume}{97}
  (\bibinfo{year}{2018}) \bibinfo{pages}{043843}.
\bibitem[{Cao et~al.(2018)Cao, Han, Chen, Jiang, Nguyen, Turgut, Fuchs, and
  Muller}]{cao2018theory}
\bibinfo{author}{M.~C. Cao}, \bibinfo{author}{Y.~Han},
  \bibinfo{author}{Z.~Chen}, \bibinfo{author}{Y.~Jiang}, \bibinfo{author}{K.~X.
  Nguyen}, \bibinfo{author}{E.~Turgut}, \bibinfo{author}{G.~D. Fuchs},
  \bibinfo{author}{D.~A. Muller},
\newblock \bibinfo{title}{Theory and practice of electron diffraction from
  single atoms and extended objects using an empad},
\newblock \bibinfo{journal}{Microscopy} \bibinfo{volume}{67}
  (\bibinfo{year}{2018}) \bibinfo{pages}{i150--i161}.
\bibitem[{Das et~al.(2019)Das, Tang, Hong, Gon{\c{c}}alves, McCarter, Klewe,
  Nguyen, G{\'o}mez-Ortiz, Shafer, Arenholz et~al.}]{das2019observation}
\bibinfo{author}{S.~Das}, \bibinfo{author}{Y.~Tang}, \bibinfo{author}{Z.~Hong},
  \bibinfo{author}{M.~Gon{\c{c}}alves}, \bibinfo{author}{M.~McCarter},
  \bibinfo{author}{C.~Klewe}, \bibinfo{author}{K.~Nguyen},
  \bibinfo{author}{F.~G{\'o}mez-Ortiz}, \bibinfo{author}{P.~Shafer},
  \bibinfo{author}{E.~Arenholz}, et~al.,
\newblock \bibinfo{title}{Observation of room-temperature polar skyrmions},
\newblock \bibinfo{journal}{Nature} \bibinfo{volume}{568}
  (\bibinfo{year}{2019}) \bibinfo{pages}{368--372}.
\bibitem[{Deb et~al.(2020)Deb, Cao, Han, Holtz, Xie, Park, Hovden, and
  Muller}]{deb2020imaging}
\bibinfo{author}{P.~Deb}, \bibinfo{author}{M.~C. Cao},
  \bibinfo{author}{Y.~Han}, \bibinfo{author}{M.~E. Holtz},
  \bibinfo{author}{S.~Xie}, \bibinfo{author}{J.~Park},
  \bibinfo{author}{R.~Hovden}, \bibinfo{author}{D.~A. Muller},
\newblock \bibinfo{title}{Imaging polarity in two dimensional materials by
  breaking friedel's law},
\newblock \bibinfo{journal}{Ultramicroscopy} \bibinfo{volume}{215}
  (\bibinfo{year}{2020}) \bibinfo{pages}{113019}.
\bibitem[{Mahr et~al.(2022)Mahr, Grieb, Krause, Schowalter, and
  Rosenauer}]{MAHR2022113503}
\bibinfo{author}{C.~Mahr}, \bibinfo{author}{T.~Grieb}, \bibinfo{author}{F.~F.
  Krause}, \bibinfo{author}{M.~Schowalter}, \bibinfo{author}{A.~Rosenauer},
\newblock \bibinfo{title}{Towards the interpretation of a shift of the central
  beam in nano-beam electron diffraction as a change in mean inner potential},
\newblock \bibinfo{journal}{Ultramicroscopy} \bibinfo{volume}{236}
  (\bibinfo{year}{2022}) \bibinfo{pages}{113503}.
\bibitem[{Nguyen et~al.(2020)Nguyen, Jiang, Cao, Purohit, Yadav,
  Garc{\'\i}a-Fern{\'a}ndez, Tate, Chang, Aguado-Puente, {\'I}{\~n}iguez
  et~al.}]{nguyen2020transferring}
\bibinfo{author}{K.~X. Nguyen}, \bibinfo{author}{Y.~Jiang},
  \bibinfo{author}{M.~C. Cao}, \bibinfo{author}{P.~Purohit},
  \bibinfo{author}{A.~K. Yadav},
  \bibinfo{author}{P.~Garc{\'\i}a-Fern{\'a}ndez}, \bibinfo{author}{M.~W. Tate},
  \bibinfo{author}{C.~S. Chang}, \bibinfo{author}{P.~Aguado-Puente},
  \bibinfo{author}{J.~{\'I}{\~n}iguez}, et~al.,
\newblock \bibinfo{title}{Transferring orbital angular momentum to an electron
  beam reveals toroidal and chiral order},
\newblock \bibinfo{journal}{arXiv preprint arXiv:2012.04134}
  (\bibinfo{year}{2020}).
\bibitem[{Brown et~al.(2018)Brown, Chen, Weyland, Ophus, Ciston, Allen, and
  Findlay}]{brown2018structure}
\bibinfo{author}{H.~G. Brown}, \bibinfo{author}{Z.~Chen},
  \bibinfo{author}{M.~Weyland}, \bibinfo{author}{C.~Ophus},
  \bibinfo{author}{J.~Ciston}, \bibinfo{author}{L.~J. Allen},
  \bibinfo{author}{S.~D. Findlay},
\newblock \bibinfo{title}{Structure retrieval at atomic resolution in the
  presence of multiple scattering of the electron probe},
\newblock \bibinfo{journal}{Physical Review Letters} \bibinfo{volume}{121}
  (\bibinfo{year}{2018}) \bibinfo{pages}{266102}.
\bibitem[{Pelz et~al.(2021)Pelz, Brown, Stonemeyer, Findlay, Zettl, Ercius,
  Zhang, Ciston, Scott, and Ophus}]{pelz2021phase}
\bibinfo{author}{P.~M. Pelz}, \bibinfo{author}{H.~G. Brown},
  \bibinfo{author}{S.~Stonemeyer}, \bibinfo{author}{S.~D. Findlay},
  \bibinfo{author}{A.~Zettl}, \bibinfo{author}{P.~Ercius},
  \bibinfo{author}{Y.~Zhang}, \bibinfo{author}{J.~Ciston},
  \bibinfo{author}{M.~Scott}, \bibinfo{author}{C.~Ophus},
\newblock \bibinfo{title}{Phase-contrast imaging of multiply-scattering
  extended objects at atomic resolution by reconstruction of the scattering
  matrix},
\newblock \bibinfo{journal}{Physical Review Research} \bibinfo{volume}{3}
  (\bibinfo{year}{2021}) \bibinfo{pages}{023159}.
\bibitem[{Findlay et~al.(2021)Findlay, Brown, Pelz, Ophus, Ciston, and
  Allen}]{findlay2021scattering}
\bibinfo{author}{S.~D. Findlay}, \bibinfo{author}{H.~G. Brown},
  \bibinfo{author}{P.~M. Pelz}, \bibinfo{author}{C.~Ophus},
  \bibinfo{author}{J.~Ciston}, \bibinfo{author}{L.~J. Allen},
\newblock \bibinfo{title}{Scattering matrix determination in crystalline
  materials from {4D} scanning transmission electron microscopy at a single
  defocus value},
\newblock \bibinfo{journal}{Microscopy and Microanalysis} \bibinfo{volume}{27}
  (\bibinfo{year}{2021}) \bibinfo{pages}{744--757}.
\bibitem[{Brown et~al.(2022)Brown, Pelz, Hsu, Zhang, Ramesh, Inzani, Sheridan,
  Griffin, Schloz, Pekin et~al.}]{brown2022three}
\bibinfo{author}{H.~G. Brown}, \bibinfo{author}{P.~M. Pelz},
  \bibinfo{author}{S.-L. Hsu}, \bibinfo{author}{Z.~Zhang},
  \bibinfo{author}{R.~Ramesh}, \bibinfo{author}{K.~Inzani},
  \bibinfo{author}{E.~Sheridan}, \bibinfo{author}{S.~M. Griffin},
  \bibinfo{author}{M.~Schloz}, \bibinfo{author}{T.~C. Pekin}, et~al.,
\newblock \bibinfo{title}{A three-dimensional reconstruction algorithm for
  scanning transmission electron microscopy data from a single sample
  orientation},
\newblock \bibinfo{journal}{Microscopy and Microanalysis} \bibinfo{volume}{28}
  (\bibinfo{year}{2022}) \bibinfo{pages}{1632--1640}.
\bibitem[{Sturkey(1962)}]{sturkey1962calculation}
\bibinfo{author}{L.~Sturkey},
\newblock \bibinfo{title}{The calculation of electron diffraction intensities},
\newblock \bibinfo{journal}{Proceedings of the Physical Society (1958-1967)}
  \bibinfo{volume}{80} (\bibinfo{year}{1962}) \bibinfo{pages}{321}.
\bibitem[{Pennington et~al.(2014)Pennington, Van~den Broek, and
  Koch}]{PhysRevB.89.205409}
\bibinfo{author}{R.~S. Pennington}, \bibinfo{author}{W.~Van~den Broek},
  \bibinfo{author}{C.~T. Koch},
\newblock \bibinfo{title}{Third-dimension information retrieval from a single
  convergent-beam transmission electron diffraction pattern using an artificial
  neural network},
\newblock \bibinfo{journal}{Phys. Rev. B} \bibinfo{volume}{89}
  (\bibinfo{year}{2014}) \bibinfo{pages}{205409}.
\bibitem[{Pennington and Koch(2015{\natexlab{a}})}]{PENNINGTON2015105}
\bibinfo{author}{R.~S. Pennington}, \bibinfo{author}{C.~T. Koch},
\newblock \bibinfo{title}{Retrieving depth-direction information from {TEM}
  diffraction data under reciprocal-space sampling variation},
\newblock \bibinfo{journal}{Ultramicroscopy} \bibinfo{volume}{148}
  (\bibinfo{year}{2015}{\natexlab{a}}) \bibinfo{pages}{105--114}.
\bibitem[{Pennington and Koch(2015{\natexlab{b}})}]{PENNINGTON201542}
\bibinfo{author}{R.~S. Pennington}, \bibinfo{author}{C.~T. Koch},
\newblock \bibinfo{title}{A three-dimensional polarization domain retrieval
  method from electron diffraction data},
\newblock \bibinfo{journal}{Ultramicroscopy} \bibinfo{volume}{155}
  (\bibinfo{year}{2015}{\natexlab{b}}) \bibinfo{pages}{42--48}.
\bibitem[{Pennington et~al.(2018)Pennington, Coll, Estrad\'e, Peir\'o, and
  Koch}]{PhysRevB.97.024112}
\bibinfo{author}{R.~S. Pennington}, \bibinfo{author}{C.~Coll},
  \bibinfo{author}{S.~Estrad\'e}, \bibinfo{author}{F.~Peir\'o},
  \bibinfo{author}{C.~T. Koch},
\newblock \bibinfo{title}{Neural-network-based depth-resolved multiscale
  structural optimization using density functional theory and electron
  diffraction data},
\newblock \bibinfo{journal}{Phys. Rev. B} \bibinfo{volume}{97}
  (\bibinfo{year}{2018}) \bibinfo{pages}{024112}.
\bibitem[{Jacob et~al.(2008)Jacob, Zuo, Lefebvre, and Cordier}]{Jacob2008}
\bibinfo{author}{D.~Jacob}, \bibinfo{author}{J.~Zuo},
  \bibinfo{author}{A.~Lefebvre}, \bibinfo{author}{Y.~Cordier},
\newblock \bibinfo{title}{Composition analysis of semiconductor quantum wells
  by energy filtered convergent-beam electron diffraction},
\newblock \bibinfo{journal}{Ultramicroscopy} \bibinfo{volume}{108}
  (\bibinfo{year}{2008}) \bibinfo{pages}{358--366}.
\bibitem[{Yadav et~al.(2016)Yadav, Nelson, Hsu, Hong, Clarkson, Schlep{\"u}tz,
  Damodaran, Shafer, Arenholz, Dedon et~al.}]{yadav2016observation}
\bibinfo{author}{A.~Yadav}, \bibinfo{author}{C.~Nelson},
  \bibinfo{author}{S.~Hsu}, \bibinfo{author}{Z.~Hong},
  \bibinfo{author}{J.~Clarkson}, \bibinfo{author}{C.~Schlep{\"u}tz},
  \bibinfo{author}{A.~Damodaran}, \bibinfo{author}{P.~Shafer},
  \bibinfo{author}{E.~Arenholz}, \bibinfo{author}{L.~Dedon}, et~al.,
\newblock \bibinfo{title}{Observation of polar vortices in oxide
  superlattices},
\newblock \bibinfo{journal}{Nature} \bibinfo{volume}{530}
  (\bibinfo{year}{2016}) \bibinfo{pages}{198--201}.
\bibitem[{Shao et~al.(2022)Shao, Schnitzer, Ruf, Gorobtsov, Dai, Goodge, Yang,
  Nair, Stoica, Freeland, Ruff, Chen, Schlom, Shen, Kourkoutis, and
  Singer}]{shao_arxiv_2022}
\bibinfo{author}{Z.~Shao}, \bibinfo{author}{N.~Schnitzer},
  \bibinfo{author}{J.~Ruf}, \bibinfo{author}{O.~Y. Gorobtsov},
  \bibinfo{author}{C.~Dai}, \bibinfo{author}{B.~H. Goodge},
  \bibinfo{author}{T.~Yang}, \bibinfo{author}{H.~Nair}, \bibinfo{author}{V.~A.
  Stoica}, \bibinfo{author}{J.~W. Freeland}, \bibinfo{author}{J.~Ruff},
  \bibinfo{author}{L.-Q. Chen}, \bibinfo{author}{D.~G. Schlom},
  \bibinfo{author}{K.~M. Shen}, \bibinfo{author}{L.~F. Kourkoutis},
  \bibinfo{author}{A.~Singer}, \bibinfo{title}{Real-space imaging of polar and
  elastic nano-textures in thin films via inversion of diffraction data},
  \bibinfo{year}{2022}.
\bibitem[{De~Graef(2003)}]{de2003introduction}
\bibinfo{author}{M.~De~Graef}, \bibinfo{title}{Introduction to conventional
  transmission electron microscopy}, \bibinfo{publisher}{Cambridge university
  press}, \bibinfo{year}{2003}.
\bibitem[{Singh et~al.(2018)Singh, Mills, and Graef}]{SINGH201832}
\bibinfo{author}{S.~Singh}, \bibinfo{author}{M.~Mills}, \bibinfo{author}{M.~D.
  Graef},
\newblock \bibinfo{title}{Dynamical scattering image simulations for two-phase
  $\gamma$--$\gamma'$ microstructures: A theoretical model},
\newblock \bibinfo{journal}{Ultramicroscopy} \bibinfo{volume}{185}
  (\bibinfo{year}{2018}) \bibinfo{pages}{32--41}.
\bibitem[{Najfeld and Havel(1995)}]{NAJFELD1995321}
\bibinfo{author}{I.~Najfeld}, \bibinfo{author}{T.~Havel},
\newblock \bibinfo{title}{Derivatives of the matrix exponential and their
  computation},
\newblock \bibinfo{journal}{Advances in Applied Mathematics}
  \bibinfo{volume}{16} (\bibinfo{year}{1995}) \bibinfo{pages}{321--375}.
\bibitem[{Weickenmeier and Kohl(1991)}]{weickenmeier1991computation}
\bibinfo{author}{A.~Weickenmeier}, \bibinfo{author}{H.~Kohl},
\newblock \bibinfo{title}{Computation of absorptive form factors for
  high-energy electron diffraction},
\newblock \bibinfo{journal}{Acta Crystallographica Section A: Foundations of
  Crystallography} \bibinfo{volume}{47} (\bibinfo{year}{1991})
  \bibinfo{pages}{590--597}.
\bibitem[{Boyd et~al.(2011)Boyd, Parikh, Chu, Peleato, Eckstein
  et~al.}]{boyd2011distributed}
\bibinfo{author}{S.~Boyd}, \bibinfo{author}{N.~Parikh},
  \bibinfo{author}{E.~Chu}, \bibinfo{author}{B.~Peleato},
  \bibinfo{author}{J.~Eckstein}, et~al.,
\newblock \bibinfo{title}{Distributed optimization and statistical learning via
  the alternating direction method of multipliers},
\newblock \bibinfo{journal}{Foundations and Trends in Machine learning}
  \bibinfo{volume}{3} (\bibinfo{year}{2011}) \bibinfo{pages}{1--122}.
\bibitem[{JM(1998)}]{jm1998quantitative}
\bibinfo{author}{Z.~JM},
\newblock \bibinfo{title}{Quantitative convergent beam electron diffraction},
\newblock \bibinfo{journal}{Materials Transactions, JIM} \bibinfo{volume}{39}
  (\bibinfo{year}{1998}) \bibinfo{pages}{938--946}.
\bibitem[{Wu et~al.(2020)Wu, Meng, and Zhu}]{WU2020113095}
\bibinfo{author}{L.~Wu}, \bibinfo{author}{Q.~Meng}, \bibinfo{author}{Y.~Zhu},
\newblock \bibinfo{title}{Mapping valence electron distributions with multipole
  density formalism using {4D-STEM}},
\newblock \bibinfo{journal}{Ultramicroscopy} \bibinfo{volume}{219}
  (\bibinfo{year}{2020}) \bibinfo{pages}{113095}.
\bibitem[{Mortensen et~al.(2005)Mortensen, Hansen, and
  Jacobsen}]{PhysRevB.71.035109}
\bibinfo{author}{J.~J. Mortensen}, \bibinfo{author}{L.~B. Hansen},
  \bibinfo{author}{K.~W. Jacobsen},
\newblock \bibinfo{title}{Real-space grid implementation of the projector
  augmented wave method},
\newblock \bibinfo{journal}{Phys. Rev. B} \bibinfo{volume}{71}
  (\bibinfo{year}{2005}) \bibinfo{pages}{035109}.
\bibitem[{Madsen and Susi(2021)}]{10.12688/openreseurope.13015.2}
\bibinfo{author}{J.~Madsen}, \bibinfo{author}{T.~Susi},
\newblock \bibinfo{title}{The {abTEM} code: transmission electron microscopy
  from first principles [version 2; peer review: 2 approved]},
\newblock \bibinfo{journal}{Open Research Europe} \bibinfo{volume}{1}
  (\bibinfo{year}{2021}).
\bibitem[{Savitzky et~al.(2021)Savitzky, Zeltmann, Hughes, Brown, Zhao, Pelz,
  Pekin, Barnard, Donohue, DaCosta et~al.}]{savitzky2021py4dstem}
\bibinfo{author}{B.~H. Savitzky}, \bibinfo{author}{S.~E. Zeltmann},
  \bibinfo{author}{L.~A. Hughes}, \bibinfo{author}{H.~G. Brown},
  \bibinfo{author}{S.~Zhao}, \bibinfo{author}{P.~M. Pelz},
  \bibinfo{author}{T.~C. Pekin}, \bibinfo{author}{E.~S. Barnard},
  \bibinfo{author}{J.~Donohue}, \bibinfo{author}{L.~R. DaCosta}, et~al.,
\newblock \bibinfo{title}{{py4DSTEM}: A software package for four-dimensional
  scanning transmission electron microscopy data analysis},
\newblock \bibinfo{journal}{Microscopy and Microanalysis} \bibinfo{volume}{27}
  (\bibinfo{year}{2021}) \bibinfo{pages}{712--743}.
\bibitem[{Ophus et~al.(2022)Ophus, Zeltmann, Bruefach, Rakowski, Savitzky,
  Minor, and Scott}]{ophus2022automated}
\bibinfo{author}{C.~Ophus}, \bibinfo{author}{S.~E. Zeltmann},
  \bibinfo{author}{A.~Bruefach}, \bibinfo{author}{A.~Rakowski},
  \bibinfo{author}{B.~H. Savitzky}, \bibinfo{author}{A.~M. Minor},
  \bibinfo{author}{M.~C. Scott},
\newblock \bibinfo{title}{Automated crystal orientation mapping in py4dstem
  using sparse correlation matching},
\newblock \bibinfo{journal}{Microscopy and Microanalysis} \bibinfo{volume}{28}
  (\bibinfo{year}{2022}) \bibinfo{pages}{390--403}.
\bibitem[{Zeltmann et~al.(2022)Zeltmann, Minor, and Ophus}]{zeltmann20224dstem}
\bibinfo{author}{S.~E. Zeltmann}, \bibinfo{author}{A.~M. Minor},
  \bibinfo{author}{C.~Ophus},
\newblock \bibinfo{title}{{4D-STEM} measurement of thickness and orientation by
  {Bloch} wave dynamical diffraction matching},
\newblock \bibinfo{journal}{Microscopy and Microanalysis} \bibinfo{volume}{28}
  (\bibinfo{year}{2022}) \bibinfo{pages}{382--383}.
\bibitem[{Susarla et~al.(2022)Susarla, Hsu, G\'{o}mez-Ortiz, Savitzky, Das,
  Behera, Junquera, Ercius, Ramesh, and Ophus}]{susarla2022emergence}
\bibinfo{author}{S.~Susarla}, \bibinfo{author}{S.~Hsu},
  \bibinfo{author}{F.~G\'{o}mez-Ortiz}, \bibinfo{author}{B.~Savitzky},
  \bibinfo{author}{S.~Das}, \bibinfo{author}{P.~Behera},
  \bibinfo{author}{J.~Junquera}, \bibinfo{author}{P.~Ercius},
  \bibinfo{author}{R.~Ramesh}, \bibinfo{author}{C.~Ophus},
\newblock \bibinfo{title}{The emergence of three-dimensional chiral domain
  walls in polar vortices},
\newblock \bibinfo{journal}{Manuscript under review}  (\bibinfo{year}{2022}).
\bibitem[{Monga et~al.(2021)Monga, Li, and Eldar}]{monga2021algorithm}
\bibinfo{author}{V.~Monga}, \bibinfo{author}{Y.~Li}, \bibinfo{author}{Y.~C.
  Eldar},
\newblock \bibinfo{title}{Algorithm unrolling: Interpretable, efficient deep
  learning for signal and image processing},
\newblock \bibinfo{journal}{IEEE Signal Processing Magazine}
  \bibinfo{volume}{38} (\bibinfo{year}{2021}) \bibinfo{pages}{18--44}.
\bibitem[{Hong et~al.(2021)Hong, Zeltmann, Savitzky, DaCosta, M{\"u}ller,
  Minor, Bustillo, and Ophus}]{hong2021multibeam}
\bibinfo{author}{X.~Hong}, \bibinfo{author}{S.~E. Zeltmann},
  \bibinfo{author}{B.~H. Savitzky}, \bibinfo{author}{L.~R. DaCosta},
  \bibinfo{author}{A.~M{\"u}ller}, \bibinfo{author}{A.~M. Minor},
  \bibinfo{author}{K.~C. Bustillo}, \bibinfo{author}{C.~Ophus},
\newblock \bibinfo{title}{Multibeam electron diffraction},
\newblock \bibinfo{journal}{Microscopy and Microanalysis} \bibinfo{volume}{27}
  (\bibinfo{year}{2021}) \bibinfo{pages}{129--139}.

\end{thebibliography}

\end{document}